4# Direct observation of time correlated single-electron tunneling

Jonas Bylander, Tim Duty, and Per Delsing*Abstract*—We report a direct detection of time correlated single-electron tunneling oscillations in a series array of small tunnel junctions. Here the current, *I*, is made up of a lattice of charge solitons moving throughout the array by time correlated tunneling with the frequency $f=I/e$, where *e* is the electron charge. To detect the single charges, we have integrated the array with a radio-frequency single-electron transistor (RF-SET) and employed two different methods to couple the array to the SET input: by direct injection through a tunnel junction, and by capacitive coupling. In this paper we report the results from the latter type of charge input, where we have observed the oscillations in the frequency domain and measured currents from 50 to 250 fA by means of electron counting.## I. INTRODUCTION

RECENTLY, we have measured a very small dc current by direct counting of the single electrons that pass by in a microelectronic circuit [1]. This may have implications for metrology as a possible new current standard, and as a new tool to directly study charge transfer in mesoscopic devices.

When electrons are forced to move in one dimension, correlations appear at low temperature due to Coulomb repulsion. In a one-dimensional array [2] of metallic islands, separated by small tunnel junctions of resistance greater than the Klitzing resistance, $R_K = h/e^2 = 26$ kΩ, the addition of a single charge to one island polarizes the neighboring islands and prevents other charges from approaching. The potential profile so formed is often called a "charge soliton" since it can move throughout the array by tunneling, without changing its form. It extends over a distance of $M=(C_A/C_0)^{1/2}$ islands, where $C_A$ is the capacitance of each junction in the array, and $C_0$ is the stray capacitance of an island. When several solitons are present inside the array, a charge lattice is formed, and above a certain threshold voltage, $V_t$, a current consisting of such single charge solitons starts to flow in the array (see Fig. 1(a) for its current-voltage characteristics). For a sufficiently small current, *I*, the space correlation of the charges results in a time correlated oscillation of the potential of any island in the array, which permits the electrons to be counted. The average frequency of these oscillations is

$$f = I/e, \quad (1)$$

where *e* is the charge of the electron.

The authors are with the Dept. of Microtechnology and Nanoscience, Chalmers University of Technology, SE-412 96 Göteborg, Sweden.
This work was supported by the Swedish SSF and VR, the Wallenberg foundation, and by the EU research project COUNT.Since this equation relates current and frequency by a fundamental constant only, the detection of the frequency provides a self-calibrated measure of the current. Previously, single-electron pumps [3] have been able to *generate* currents given by the relation $I=ef$ with very good accuracy. There, one electron at a time is "clocked" through a circuit by applying synchronized external signals of the frequency *f* to gates in the circuit. Conversely, the counting of every electron passing through the array as in (1) is a current *measurement*. By coupling many such counters in parallel, it would be possible to measure higher currents than what can be generated with single-electron pumps.

## II. CHARGE DETECTION

In order to detect the oscillations (1) we have used a single-electron transistor (SET) [4]–[5] embedded in an *LC* tank circuit and operated in the radio-frequency mode (RF-SET) [6], allowing fast and very sensitive readout. The charge

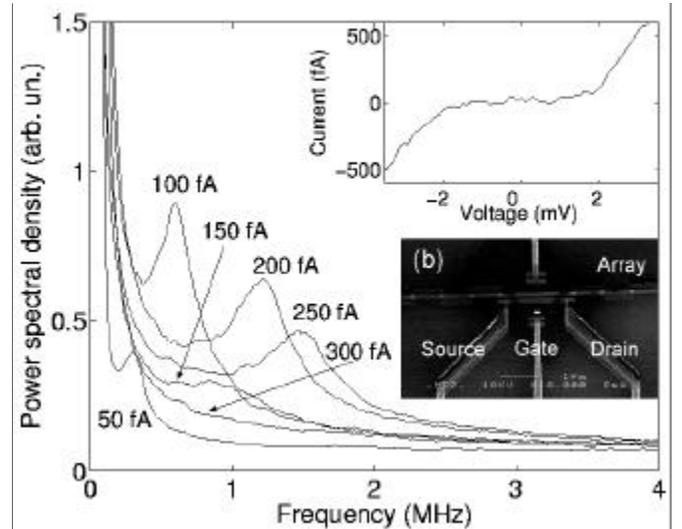

Fig. 1. Main graph. Power spectral densities of the reflected power from the SET as measured with a spectrum analyzer with 30 kHz resolution bandwidth. The periodic passage of electrons in the array modulates the SET conductance with a frequency that corresponds to the current as $f=I/e$. The array currents, *I*, are indicated in the graph and the corresponding frequencies are 0.31, 0.62, 0.94, 1.2, 1.6, and 1.9 MHz, respectively. The lower SNR of the 150 fA peak is likely due to the read-out SET being detuned from its working point. The large low-frequency tail is due to the spectrum analyzer where the swept local oscillator leaks through the internal mixer. The data were taken at $B_\parallel=0.45$ T.

Insets. (a) Current–voltage characteristics of the superconducting array ($B_\parallel=0.475$ T) showing the Coulomb gap and the gradual onset of current above the threshold, $V_t$. (b) Scanning electron micrograph of the device showing the array, the SET and two (dc) tuning gates.

changes on the SET modulate the reflected power of the RF signal launched towards the tank circuit.

The charge input to the SET can be accomplished in two different ways: either by injecting the current directly into the SET island through a tunnel junction (array-coupled SET) as in [1], or in the conventional way by a capacitive coupling between one of the array islands and the SET island. The former gives a good signal-to-noise ratio (SNR) since the full charge of the electron is injected into the SET island. However, in this way the amplifier becomes highly non-linear as the output signal of the SET is $1e$-periodic in the gate charge and the injected charge moves the amplifier far away from its working point. Moreover, as a current of at least several pA is flowing through the SET, the array-coupled SET is bound to exert a greater backaction than the capacitively coupled version. Also backtunneling of electrons from the SET into the array may take place in the array-coupled device, thus reducing the accuracy. With the capacitive coupling, on the other hand, the SET senses only a fraction of the charge on the coupling capacitor, $C_C$, and consequently the SNR is significantly lower. The SET can then be operated in the linear regime, however. In this paper we report measurements on a device, which is capacitively coupled.

## III. Experimental Setup

We performed the measurements at $T=30$ mK in a dilution refrigerator using strongly filtered cables from the room temperature electronics. We applied a stable bias to the array either by using a transimpedance amplifier or a current source. In all cases, the array was voltage biased due to the cabling capacitances.

We fabricated the samples with electron beam lithography and three-angle shadow evaporation to deposit aluminum and achieve different oxide thickness for the SET and array tunnel junctions, see Fig. 1(b). The SET resistance was $R^{SET}=30$ kΩ, and the charging energy $E_C^{SET}/k_B=1.6$ K. We obtained a charge sensitivity of $dq=2\cdot10^{-5}$ $e/Hz^{1/2}$ and a measurement bandwidth of $\Delta f=10$ MHz at an RF carrier frequency of $f_0=372$ MHz. The array had $N=50$ junctions, each of resistance $R_N^{Array}=940$ kΩ; the junction and stray capacitances were $C_A=0.42$ fF and $C_0=0.030$ fF, respectively, giving the charging energy $E_C^{Array}/k_B=2.2$ K per junction and a soliton length of $M=(C_A/C_0)^{1/2}=3.7$ islands (i.e., $N>>M$). We estimate the capacitive coupling between the middle island of the array and the SET island to be $C_C/C_\Sigma \sim 10$ %, where $C_C=0.05$ fF is the coupling capacitance and $C_\Sigma=0.6$ fF is the total capacitance of the SET island.

## IV. Results and Discussion

We have recently reported time and frequency domain data for the array-coupled counter elsewhere [1]. There, we showed a primary current measurement by electron counting in the range from 5 fA to 1 pA, and the SNR of the peak in the power spectrum of the reflected signal was approximately 8 dB at the best. Here we report the first results for a counter with a capacitively coupled SET. In Fig. 1 the time-correlation of the single-electron tunneling oscillations shows as a peak in the averaged frequency domain data of the reflected signal from the RF-SET; the SNR was here substantially lower, only about 1 to 2 dB.

As in [1], this experiment was performed in the superconducting state, at a magnetic flux density (applied parallel to the film of the sample) of $B_\parallel=0.45$ T, where the critical flux density is $B_{\parallel,C}=0.65$ T. Here we take advantage of the high subgap resistance (estimated to be 100 times $R_N^{Array}$) that lowers the sensitivity to array bias voltage fluctuations. However, the high field effectively suppresses the superconducting gap, $\Delta/k_B=0.6$ K, and furthermore the Josephson coupling energy is very small in our junctions: $E_J/k_B<10$ mK$<<T$. Therefore we expect to see only quasi-particles and no Cooper pairs at this field although every island in the array is superconducting.

In order to improve the accuracy of our electron counter, we believe that a read-out scheme with a "differential" SET [7] and cross-correlated detection would increase the SNR. Then two (or more) SETs could be coupled capacitively to different islands in the array. If two SETs, tuned to opposite slopes of their transfer functions, both share the same $LC$ circuit, then one of them gives a maximum signal when a charge is present on its adjacent array island, whereas the other one gives a maximum signal in the absence of a charge on its corresponding island. In this way the two signals add constructively, but any uncorrelated or common mode noise does not, thus improving the SNR. Alternatively, multiplexed SETs could be used [8].

## V. Conclusion

To conclude, we have measured very small currents in a one-dimensional array of small tunnel junctions by counting the single electrons with a capacitively coupled SET. This device has potentially lower backaction on the current flow in the array than the array-coupled SET [1], but substantially lower signal-to-noise ratio.

### Acknowledgment

We thank K. Bladh, D. Gunnarsson, S. Kafanov and M. Taslakov for technical assistance.